\documentclass[arps,prd,onecolumn,showpacs,preprintnumbers]{revtex4}
\usepackage{graphicx}
\usepackage{amsmath}
\usepackage{amsfonts}
\usepackage{amssymb}
\usepackage{color}
\usepackage{epsf}

\newcommand{\be}{\begin{equation}}         
\newcommand{\ee}{\end{equation}}
\newcommand{\ba}{\begin{eqnarray}}
\newcommand{\ea}{\end{eqnarray}}
\newcommand{\nn}{\nonumber}

\newcommand{\Mpc}{{\rm Mpc}}
\newcommand{\Gpc}{{\rm Gpc}}
\newcommand{\eV}{{\rm eV}}

\newcommand\lsim{\mathrel{\rlap{\lower4pt\hbox{\hskip1pt$\sim$}}
        \raise1pt\hbox{$<$}}}
\newcommand\gsim{\mathrel{\rlap{\lower4pt\hbox{\hskip1pt$\sim$}}
        \raise1pt\hbox{$>$}}}
\def\k{{\bf k}}
\def\x{{\bf x}}
\def\thetaB{\mbox{\boldmath$\theta$}}

\begin{document}
\preprint{YITP-SB-16-17}

\title{Neutrino mass without cosmic variance}
\author{ Marilena LoVerde}
\affiliation{C. N. Yang Institute for Theoretical Physics, Department of Physics \& Astronomy, Stony Brook University, Stony Brook, NY, 11794, U.S.A.}

\begin{abstract}
Measuring the absolute scale of the neutrino masses is one of the most exciting opportunities available with near-term cosmological data sets. Two quantities that are sensitive to neutrino mass, scale-dependent halo bias $b(k)$ and the linear growth parameter $f(k)$ inferred from redshift-space distortions, can be measured without cosmic variance. Unlike the amplitude of the matter power spectrum, which always has a finite error, the error on $b(k)$ and $f(k)$ continues to decrease as the number density of tracers increases. This paper presents forecasts for statistics of galaxy and lensing fields that are sensitive to neutrino mass via $b(k)$ and $f(k)$. The constraints on neutrino mass from the auto- and cross-power spectra of spectroscopic and photometric galaxy samples are weakened by scale-dependent bias unless a very high density of tracers is available. In the high-density limit, using multiple tracers allows cosmic variance to be beaten, and the forecasted errors on neutrino mass shrink dramatically. In practice, beating the cosmic-variance errors on neutrino mass with $b(k)$ will be a challenge, but this signal is nevertheless a new probe of neutrino effects on structure formation that is interesting in its own right. 
\end{abstract}
\maketitle

\section{introduction}
\label{sec:intro}
Cosmic neutrinos are the second most abundant particle in the Universe but their masses and their contribution to the current cosmic energy budget are not known. The neutrino contribution to the early-Universe radiation density has been detected at high significance and is consistent with three neutrinos each with number density $\bar{n}_\nu\approx 112/\rm{cm}^3$ and temperature slightly cooler than the cosmic microwave background (CMB), $T_\nu \approx 1.6 \times 10^{-4}$eV\cite{Ade:2015xua}. Today, the energy density of neutrinos is dominated by their rest mass, $\rho_\nu \approx \sum m_{\nu} \bar{n}_\nu$. Neutrino oscillation data specify the square of two mass splittings, $\Delta m_{12}^2 = 7.54 \times 10^{-5}$eV, $|\Delta m_{13}^2|\approx 2.4 \times 10^{-3}$eV \cite{Agashe:2014kda}, but not the individual masses. For $\Delta m_{13} > 0$, we have $m_{\nu 1} \gsim 0$eV, $m_{\nu 2} \gsim 0.0087$eV, $m_{\nu 3} \gsim 0.049$eV, the ``normal hierarchy.'' Whereas for $\Delta m_{13} <0$, we have $m_{\nu 1} \gsim  0.049$eV, $m_{\nu 2} \gsim 0.05$eV, and $m_{\nu 3} \gsim 0$eV, which is called the ``inverted hierarchy." If any one of the neutrino mass states is $\gsim 0.1eV$, then the oscillation data require $m_{\nu 1}\approx m_{\nu 2}\approx m_{\nu 3}$, and the hierarchy is quasidegenerate.  The oscillation data, in combination with the relic neutrino number density, therefore gives a lower limit on the neutrino contribution to the cosmic energy budget of $\Omega_\nu \equiv \rho_\nu/\rho_{critical} \gsim 0.001$. The current upper limit on $M_\nu \equiv \sum_i m_{\nu i} $, and therefore $\Omega_\nu$, is $M_\nu \lsim 0.12 - 0.5$eV  at $95\%$ confidence depending on the data set \cite{Ade:2015xua, Palanque-Delabrouille:2015pga, Cuesta:2015iho}. 

Future large-scale structure data sets such as those from the Dark Energy Spectroscopic Instrument (DESI) \cite{Levi:2013gra}, Euclid \cite{Euclid}, the Large Synoptic Survey Telescope (LSST) \cite{LSST}, Wide-Field InfraRed Survey Telescope (WFIRST) \cite{WFIRST}, SPHEREx \cite{Dore:2014cca}, and lensing measurements from Advanced ACTPol \cite{Henderson:2015nzj}, SPT-3G \cite{Benson:2014qhw}, and a Stage IV CMB experiment will have the statistical power to detect neutrino mass at the $3-4\sigma$ level  (see e.g. \cite{Abazajian:2013oma, Allison:2015qca,Manzotti:2015ozr,DiValentino:2015sam}), potentially ruling out the inverted mass hierarchy. 

For fixed background cosmology, increasing $M_\nu$ increases the fraction of the matter density in massive neutrinos, thereby suppressing the linear growth of matter perturbations with wavelength $2\pi/k < 2\pi/k_{fs}$ where $k_{fs}$ is the wavenumber corresponding to the neutrino free-streaming scale, $k_{fs} \sim m_\nu a H/T_\nu$ (for a review see \cite{Lesgourgues:2006nd}). The net suppression in the matter power spectrum, probed through galaxy clustering, weak lensing shear, or CMB lensing, is a classic cosmological test of neutrino mass \cite{Hu:1997mj}. The neutrino-induced changes to the growth of linear perturbations can also be detected via changes to the amplitude of redshift-space distortions (RSD), quantified by $f \equiv d\ln \delta_m/d\ln a$, and the scale-dependent of the linear halo bias $b$ (see e.g. \cite{Font-Ribera:2013rwa, LoVerde:2014pxa, Upadhye:2015lia}). Unlike the matter power spectrum, $b$ and $f$ do not depend on the particular realization of the density field and therefore are not subject to cosmic variance \cite{Bernstein:2011ju, McDonald:2008sh, Seljak:2008xr}. The focus of this paper is to study how these cosmic-variance free quantities can be leveraged to achieve a cosmic-variance-less measurement of the neutrino mass scale $M_\nu$. 

In \S \ref{sec:nubiasandRSD} we review the effects of neutrino mass on halo bias and redshift-space distortions, discuss how these quantities can be measured without cosmic variance, and provide estimates for survey requirements.  In \S \ref{sec:data} expressions for the observables (the 3D power spectra and the angular galaxy and lensing spectra) are presented. The assumed cosmological model, galaxy populations, and survey configurations used in the forecasts are discussed in \S\ref{sec:forecast} and the resulting forecasted constraints are presented in \S\ref{sec:results}. Conclusions and a discussion of the forecasts in the context of near-term surveys are presented in \S \ref{sec:conclusions}. 

\section{Neutrino effects beyond the matter power spectrum }
\label{sec:nubiasandRSD}
\subsection{Scale-dependent bias}
\label{ssec:neutrinobias}
\begin{figure}[t]
\begin{center}
$\begin{array}{cc}
 \includegraphics[width=0.5\textwidth]{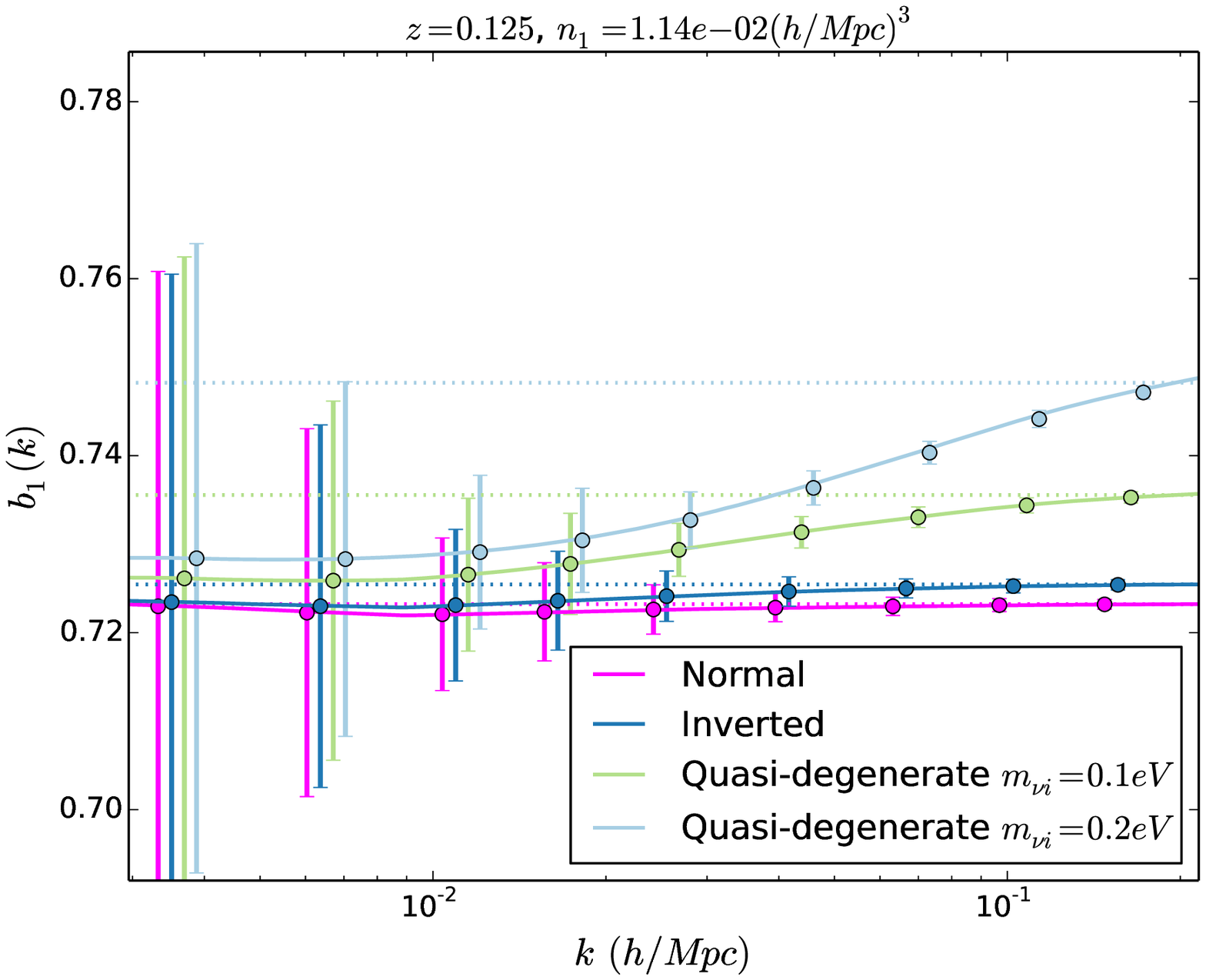} &  \includegraphics[width=0.5\textwidth]{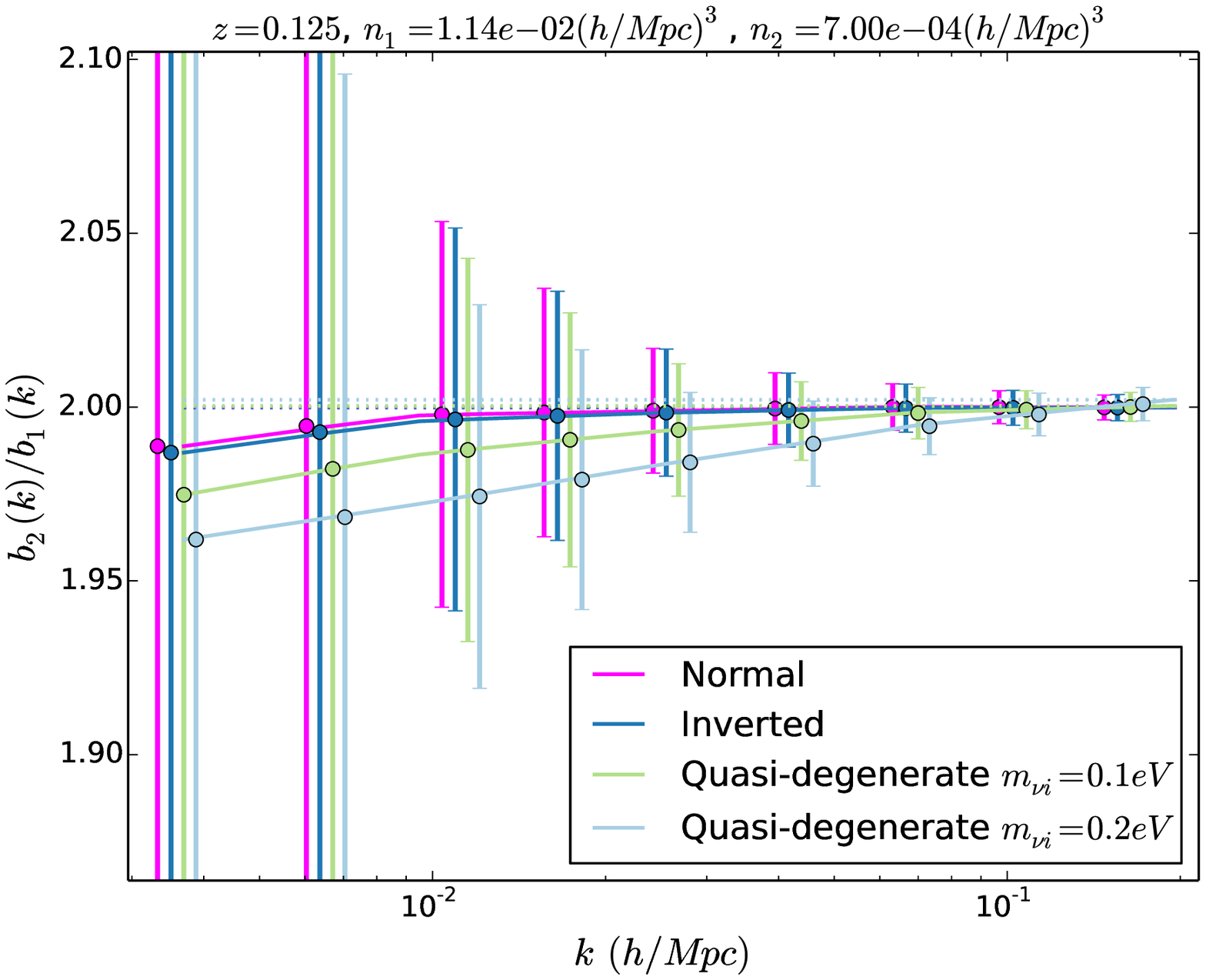}\\
 \mbox{(a)} & \mbox{(b)}
\end{array}$
 \caption{\label{fig:bofk} Left: Solid lines show the scale-dependent bias for the minimal mass normal ($m_{\nu 1}=0$eV, $m_{\nu 2} = 0.01$eV, and $m_{\nu 3} = 0.05$eV) and inverted ($m_{\nu 1} = m_{\nu 2} = 0.05$eV, $m_{\nu 3} = 0$eV) hierarchies, and two quasi-degenerate hierarchies (with $m_{\nu i} =0.1$, $0.2$ eV). For comparison, we have plotted constant biases that match $b(k)$ at at small scales (dotted lines). Error bars are the Cramer-Rao limits from Eq.~(\ref{eq:error_bh}), which assume $\hat\delta_m$ is measured in addition to $\hat\delta_g$. The number density of galaxies is $\bar{n} = 1.14 \times 10^{-2} (h/\Mpc)^3$ across a volume $\sim 0.8 h^{-3} \Gpc^3$. Right: The ratio of the scale-dependent halo biases $b_2(k)/b_1(k)$ for two populations with biases $b_1 \sim 0.7$ and $b_2 \sim 1.4$. Error bars again assume a survey volume of $\sim 0.8 h^{-3} \Gpc^3$ and that the number densities of the two populations are $\bar{n}_1 = 1.14\times 10^{-2} (h/\Mpc)^3$ and  $\bar{n}_2 = 7.0 \times 10^{-4} (h/\Mpc)^3$. The number density and bias factors are comparable to expectations for SPHEREx \cite{Dore:2014cca}. For fixed values of $b_1$ and $b_2$, the scale-dependent signals plotted here do not vary significantly with redshift.}
\end{center}
\end{figure}

\begin{figure}[t]
\begin{center}
$\begin{array}{cc}
\includegraphics[width=0.5\textwidth]{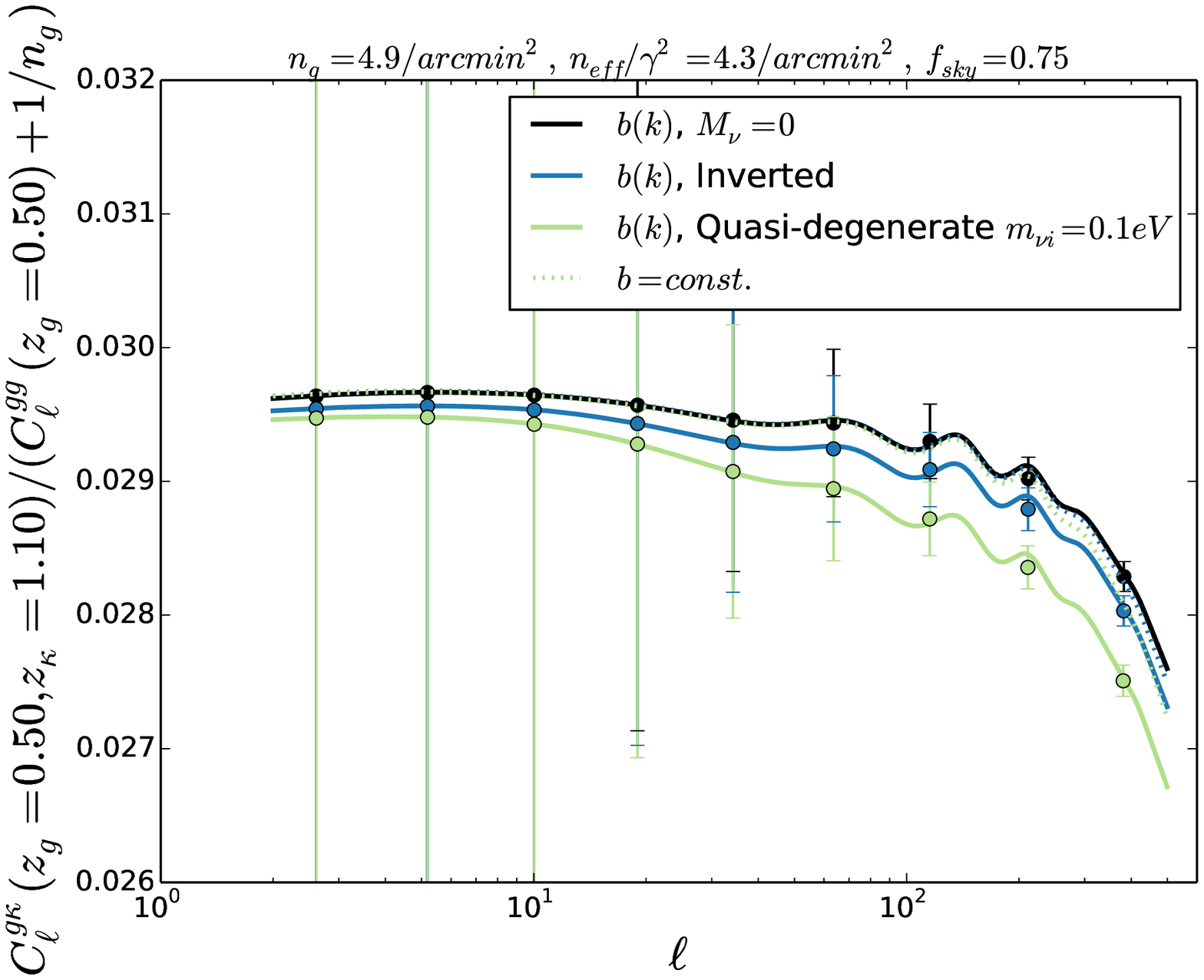} &  \includegraphics[width=0.5\textwidth]{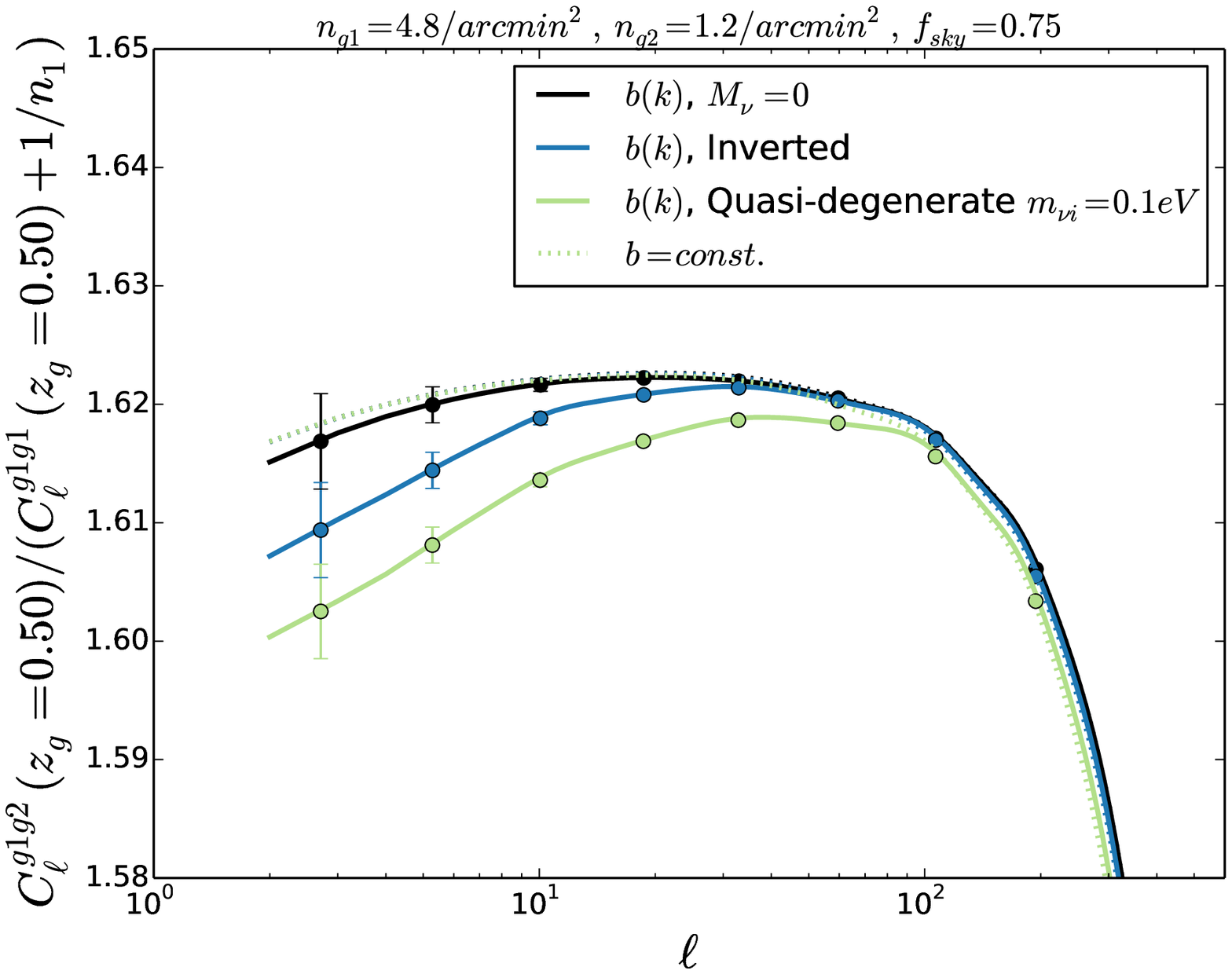} \\
\mbox{(a)} & \mbox{(b)}
\end{array}$
 \caption{\label{fig:C12OC11} Direct tests of scale dependent bias from angular power spectra in cosmologies with $M_\nu \neq 0$. Plotted is $M_\nu =0$, the minimal mass inverted hierarchy ($m_{\nu 1} = m_{\nu 2} =0.05$eV, $m_{\nu 3} =0$eV), and a quasi-degenerate hierarchy with $m_{\nu i} =0.1$eV. Left: The galaxy convergence cross-power spectrum divided by the galaxy auto-power spectrum $C_\ell^{g\kappa}/C_\ell^{gg}$ for galaxies in a redshift bin of width $\Delta z = 0.2$ at $\bar{z}_{gals} = 0.5$ and the convergence field from source galaxies also in a bin of width $\Delta z = 0.2$ at $\bar{z}_{sources} = 1.1$. Right: The cross power spectrum between two galaxy populations with the same redshift distribution ($\bar{z}_{gals} = 0.5$, $\Delta z = 0.2$) and different bias factors. In both panels, the error bars are taken to be the Cramer-Rao limit $\sigma_{C^{12}/C^{11}} = ((C^{11}_\ell+ s_1)(C_\ell^{22} + s_2) - (C_\ell^{12})^2)/(C_\ell^{11}((C_\ell^{12})^2 + (C_\ell^{11} + s_1) (C_\ell^{22} + s_2)))/\sqrt{f_{sky}(2\ell + 1)}$ where $1,2$ indicate either galaxy population or convergence field and $s_i = 1/\bar{n}$ for the galaxies and $\gamma^2/\bar{n}$ for the convergence field. The turnover at high $\ell$ is due to $s_i$ becoming larger than the auto-power spectra in the denominator in both panels. The number density of galaxies and lensing sources is comparable to what is expected from LSST. In both panels, the plotted quantities would not depend on $M_\nu$ at all if halo bias were constant. }
\end{center}
\end{figure}

\begin{figure}[t]
\begin{center}
$\begin{array}{cc}
  \includegraphics[width=0.5\textwidth]{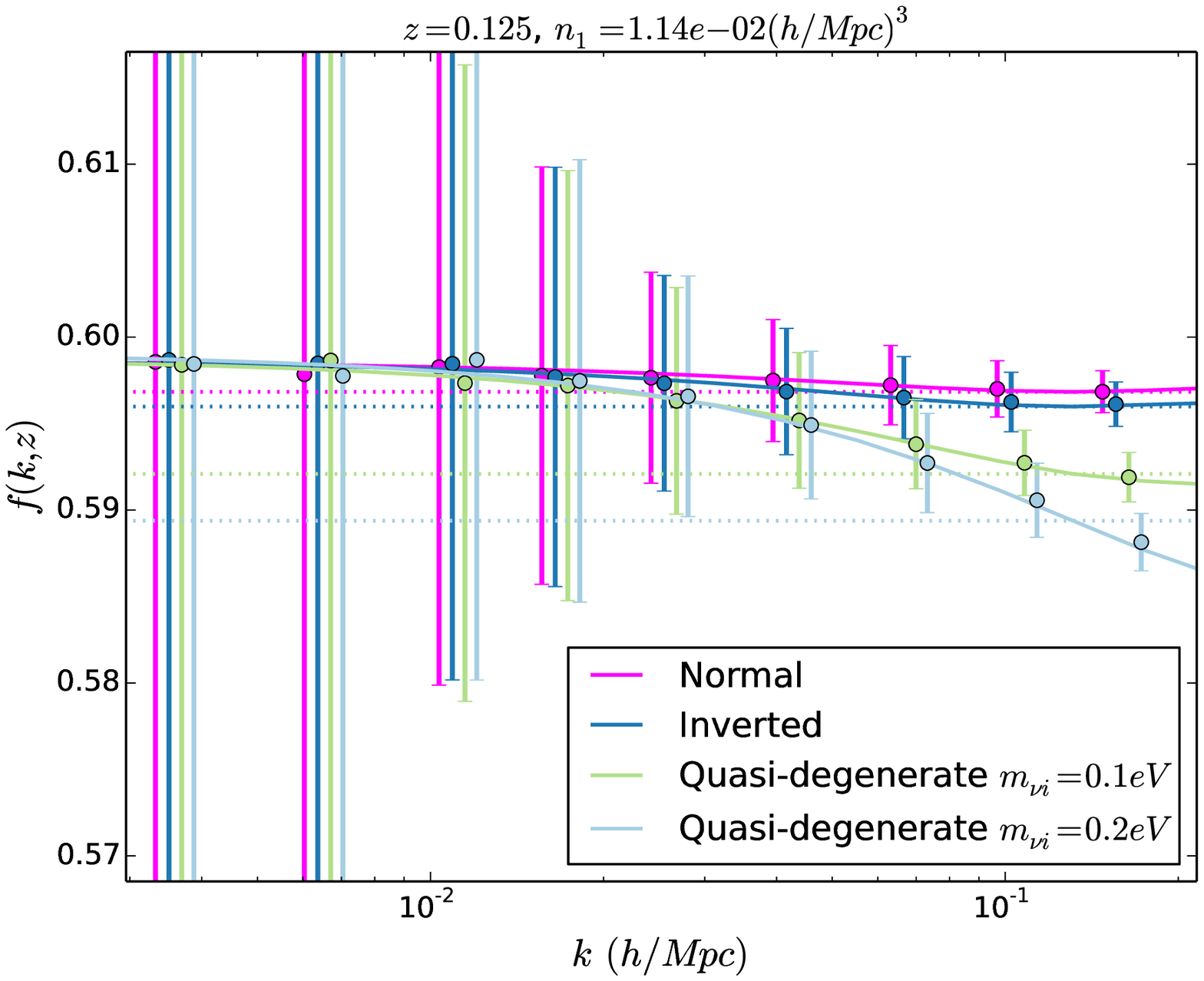} &  \includegraphics[width=0.5\textwidth]{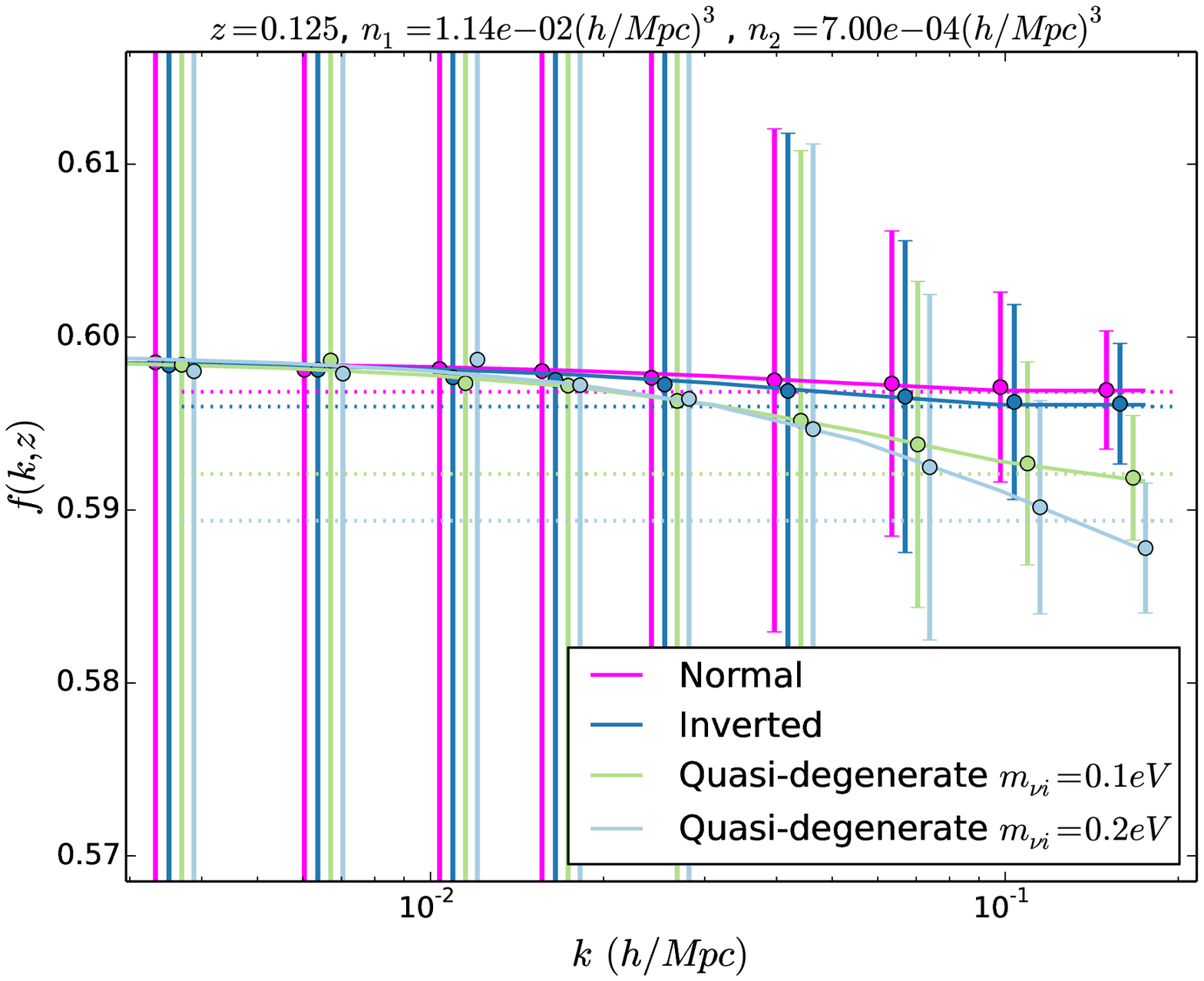}\\
   \mbox{(a)} & \mbox{(b)}
\end{array}$
 \caption{\label{fig:fofk} Left: The derivative of the linear growth factor for the minimal mass normal ($m_{\nu 1}=0$eV, $m_{\nu 2} = 0.01$eV, and $m_{\nu 3} = 0.05$eV) and inverted ($m_{\nu 1} = m_{\nu 2} = 0.05$eV, $m_{\nu 3} = 0$eV) hierarchies, and two quasi-degenerate hierarchies (with $m_{\nu i} =0.1$, $0.2$ eV). To guide the eye, we have plotted constant values of $f$ that match the $f(k)$ at at small scales (dotted lines). Error bars are the Cramer-Rao limits from \ref{ssec:RSD} which assume $\hat\delta_m$ is measured in addition to $\hat\delta_g$.  The number density of galaxies is $\bar{n} = 1.14 \times 10^{-2} (h/\Mpc)^3$ across a volume $\sim 0.8 h^{-3} \Gpc^3$. Right: The derivative of the linear growth factor as determined by two tracer populations with biases $b_1 \sim 0.7$ and $b_2 \sim 1.4$. Error bars again assume a survey volume of $\sim 0.8 h^{-3} \Gpc^3$ and that the number densities of the two populations are $\bar{n}_1 = 1.14\times 10^{-2} (h/\Mpc)^3$ and  $\bar{n}_2 = 7.0\times 10^{-4} (h/\Mpc)^3$. The number density and bias factors are comparable to expectations for SPHEREx \cite{Dore:2014cca}.}
\end{center}
\end{figure}

Massive neutrinos introduce a scale dependent feature into the halo bias $b$. We define the halo bias by 
\be
\label{eq:deltah}
\hat{\delta}_{h} = b \hat\delta_m
\ee 
where $\hat\delta_h = \hat{\delta n}_h/n_h$ is the spatial fluctuation in the number density of halos and $\hat{\delta}_m = \hat{\delta\rho}_m/\bar{\rho}_m$ is the fluctuation in the total matter (baryon, cold dark matter, and neutrino) density, and we use $\hat{}$ to indicate random quantities. Throughout we treat cold dark matter (CDM) and baryons as a single fluid with energy density $\rho_{cb} = \rho_c + \rho_b$ so that  $\rho_m = \rho_{cb} + \rho_\nu$.  

The feature introduced by massive neutrinos is a broad step around the neutrino free streaming scale and the amplitude of the step depends strongly on the neutrino mass fraction (see Figure \ref{fig:bofk}) \cite{LoVerde:2014pxa}.  The scale-dependence arises from two effects: (i) the scale-dependent growth of CDM density perturbations \cite{Hui:2007zh, Parfrey:2010uy, LoVerde:2014pxa} and (ii) the fact that halos trace CDM fluctuations rather than total (CDM + neutrino) matter density fluctuations \cite{Villaescusa-Navarro:2013pva,Castorina:2013wga,LoVerde:2014pxa}. The net scale dependence of the halo bias is given by 
\be
\label{eq:bk}
b(k) \sim \left\{\begin{array}{cc}  b_0 -  q   (b_0 -1)f_\nu  & k \ll k_{fs} \\  b_0  + b_0 f_\nu & k\gg k_{fs} \end{array}\right.
\ee
where $q \approx 0.6$,  $b_0$ is the bias factor for the halos when $m_\nu =0$, $f_\nu = \Omega_\nu/\Omega_m$. For a fixed value of $b_0$, the scale-dependent bias is roughly constant with redshift. On the other hand, for a given population of galaxies $b_0$ may vary with redshift causing the amplitude of the scale dependence to vary with redshift for that population. Note, this calculation assumes that nonlinear clustering of neutrinos in the halos can be neglected, which should be a very good approximation for $m_{\nu i} \lsim 0.1\eV$ \cite{Ringwald:2004np, LoVerde:2013lta,LoVerde:2014rxa}. 

Halo bias is particularly interesting because it is a quantity that is not subject to cosmic variance. Heuristically this can be understood as follows. A deterministic halo bias like that in Eq.~(\ref{eq:deltah}) maps one random field (the fluctuation in the matter density $\hat{\delta}_m(\x)$) to another (the fluctuation in the halo number density $\hat{\delta}_h(\x)$).  The fluctuation amplitudes $\hat\delta_h$, $\hat\delta_m$ themselves are random quantities so the cosmological information is extracted from measurements of their power spectra (or real space correlation functions). The halo-halo and matter-matter autopower spectra are
\be
\hat{P}_{hh}(k) = \frac{1}{N_{k}} \sum_{k_i \in k}\left|\hat\delta_h(k_i)\right|^2\, ,\, \hat{P}_{mm}(k) =  \frac{1}{N_k}\sum_{k_i \in k}\left|\hat\delta_m(k_i)\right|^2\,,
\ee
where $k$ is a bin with $N_{k}$ Fourier modes. For a survey with volume $V$, $N_{k} = V(4\pi)/(2\pi)^3 \Delta \ln k k^3$. There is a fundamental limit on the precision of power spectra measurements coming from cosmic variance
\be
\sigma_{\hat{P}_{hh}} = \sqrt{\frac{2}{N_k}} \left(P_{hh}+ s_h\right) \,, \quad \sigma_{\hat{P}_{mm}} = \sqrt{\frac{2}{N_k}} P_{mm} \,
\ee
where $s_h$ is the stochasticity in the halo field and even if $s_h\rightarrow 0$, the errors remain finite due to cosmic variance. On the other hand, the bias factor $b_h$ in Eq.~(\ref{eq:deltah}) is {\em not} random and can be measured even with a single Fourier mode $b_h\sim \hat\delta_h(\k)/\hat\delta_m(\k)$. The Cramer-Rao bound on the bias from measurements of the halo and matter fields is 
\be
\label{eq:error_bh}
\sigma_{b_h(k)}^2 = \frac{s_h}{N_{k}P_{mm}(k)}
\ee
where from Eq.~(\ref{eq:error_bh}) it is clear that as $s_h\rightarrow 0$, $\sigma_{b_{h}(k)} \rightarrow 0$ for any number of Fourier modes $N_k$. 

Throughout this paper we will make the standard assumption that the stochasticity term in the halo field with respect to the linear density field is Poisson shot noise $s_h = 1/\bar{n}_h$, where $\bar{n}_h$ is the number density of halos, and that there is no stochasticity between halo populations with different biases. The true stochasticity between the halo and galaxy fields may well be more complicated (see e.g. \cite{Baldauf:2013hka}) but $s= 1/\bar{n}$ is a reasonable estimate and for fixed values of the halo bias $b$ the functional dependence of our forecasts $s$ can be obtained by replacing $1/\bar{n}$ with $s$ anyway. On large scales, there are upper limits on the stochasticity between different galaxy populations and the level of stochasticity may depend on how galaxies are selected \cite{Patej:2015lwa, Pujol:2015wna}. Modifying the analysis here to account for stochasticity between populations is straightforward \cite{Seljak:2008xr}, but this is certainly an area worthy of future study. 

If instead one has two galaxy populations $1$ and $2$ with linear bias factors $b_1$, $b_2$ and stochasticities $s_1$, $s_2$ the ratio of the two bias factors $b_2/b_1$can be measured without sample variance, 
\be
\label{eq:error_b1b2}
\sigma_{b_2/b_1(k)}^2 = \frac{s_1s_2 + P_{11} \left(s_1 (b_2/b_1)^2 + s_2\right)}{P_{11}^2 N_k}\,.
\ee
The scale-dependent bias $b(k)$ and the scale-dependent ratio of the biases of two different populations with the error bars quoted in Eq.~(\ref{eq:error_bh}) and Eq.~(\ref{eq:error_b1b2}) are shown in Fig.~\ref{fig:bofk}. The scale-dependent bias can also be detected through its effects on angular power spectra of the galaxy and lensing fields (discussed further in \S \ref{sec:data}) and this is shown in Fig. ~\ref{fig:C12OC11}.

\subsection{Redshift space distortions}
\label{ssec:RSD}
With a spectroscopic survey one may measure the 3D galaxy distribution. In the presence of redshift-space distortions, the observed halo field is 
\be
\hat\delta_h(\x, \bar{z}) = \hat\delta_h(\x, \bar{z})  - \frac{1+\bar{z}}{H(\bar{z})}\frac{\partial \hat{v}_{||}}{\partial x_{||}}(\x, \bar{z})
\ee
where $v_{||}$ is the velocity along line of sight and $x_{||}$ is the line of sight distance. The continuity equation allows us to write ${\bf v}$ in terms of $\partial\hat\delta_m/\partial t$ so that the observed Fourier space galaxy fluctuations are given by
\be
\hat\delta_h(\k, \bar{z}) = \left(b_h(k, \bar{z}) + f(k,\bar{z}) \frac{k_{||}^2}{k^2}\right)\hat\delta_m(\k, \bar{z})\,. 
\ee
where we've defined
\be
f(k,z) = -(1+\bar{z})\frac{d \ln T_m(k, \bar{z})}{dz} 
\ee
where $T_{m}(k,z)$ is the matter transfer function and $k_{||}$ the line-of-sight wavenumber. Note that we have used the derivative of the $k$-dependent matter transfer function $d\ln T_{m}(k,z)/dz$ (rather than the linear growth factor $d\ln D(z)/dz$) because the linear growth is $k$-dependent in a cosmology with massive neutrinos. For $k \ll k_{fs}$, $f(k,z) \approx \Omega_m^{6/11}(z)$ and for $k \gg k_{fs}$, $f(k,z) \approx (1-\frac{3}{5}f_\nu)\Omega_m^{6/11}(z) $ \cite{Wang:1998gt,Hu:1997vi}.

 As with the halo bias discussed in \S \ref{ssec:neutrinobias}, the factor $f(k,z)$ in the redshift space distortion term is not a random quantity and is also not fundamentally limited by cosmic variance, but by the stochasticity of galaxies with respect to the density field \cite{McDonald:2008sh}. The Cramer-Rao bound on $f(k,z)$ measured from $\hat\delta_h$ and $\hat\delta_m$ and marginalizing over $b(k)$, is 

\be
\label{eq:error_fk}
\sigma_{f(k)}^2 = \frac{s}{ \frac{N_k}{2} P_{mm}(k,z)(\int d\mu \mu^4 -(\int d\mu \mu^2)^2/\int d\mu)}
\ee
where $k_{||} = \mu k$ and we've continued to use $N_{k} = V(4\pi)/(2\pi)^3 \Delta \ln k k^3$.  Note that the redshift space distortion factor introduces anisotropy into the power spectrum that allows $f(k)$ to be determined from the anisotropic power spectrum of a single tracer (as opposed to the auto- and cross-power spectra of multiple tracers or a single tracer and the underlying matter field). The Cramer-Rao bound on $f(k)$ measured from a single tracer is
\be
\sigma_{f(k), \textrm{single tracer}} = \frac{1}{\sqrt{N_k/2 \int d\mu \frac{\mu^4(b+ f\mu^2)^2}{((b+f\mu^2)^2 + s/P_{mm})^2}}}
\ee
which is finite (but independent of $P_{mm}(k)$) even in the $s\rightarrow 0$ limit. The parameter $f(k,z)$ for different neutrino mass hierarchies along with the error bars given in Eq.~(\ref{eq:error_fk}) is plotted in Fig.~\ref{fig:fofk}. Also plotted is $f(k,z)$ with the Cramer-Rao limit error bars assuming that $f(k,z)$ is determined from measuring the auto- and cross-power spectra of two galaxy populations. The expression for the Cramer-Rao bound on $f(k,z)$ from two galaxy populations is easy to calculate, but sufficiently complicated that we have not reproduced it here. Of course, in the limit that one population is very densely sampled ($1/(\bar{n}_1P_{11})\rightarrow 0$) it is just given by Eq.~(\ref{eq:error_fk}).

\subsection{Estimates of Survey Requirements}
\label{ssec:estimates}
Before proceeding it is useful to do an order of magnitude estimate of the number density of sources $\bar{n}$ and the number of Fourier modes needed to resolve the neutrino effects. We emphasize that while the amplitude of the halo bias $b_0$ also contains information about neutrino mass because halo bias is sensitive to $\sigma(M)$, it can not be treated as signal without a robust model of galaxy bias. Instead, $b_0$ is treated as an nuisance parameter in cosmological analyses (e.g. \cite{Anderson:2013yy}). The existence of a scale-dependent feature in the halo bias, however, allows for the possibility that cosmological information can be learned from the feature alone. 

With estimates of both the mass density and the galaxy field one can attempt to measure the galaxy bias $b(k)$. In this case, the magnitude of the neutrino feature we would like to resolve is 
\be
 \Delta b(k)  = b(k_{small}) - b(k_{large}) \sim b_0 f_\nu \left(1 + q \frac{b_0 -1}{b_0}\right)\,,
 \ee
 where $k_{small} > k_{fs}$ and $k_{large} < k_{fs}$. Note that the amplitude of the bias at high $k$ is generically measured more precisely than at low $k$ (e.g. for equal binning in $\ln k$, $N_{k_{small}} \gg N_{k_{large}}$) so the error on $\Delta b(k)$ is dominated by $\sigma_{b(k_{large})}$. From Eq.~(\ref{eq:error_bh}) the signal-to-noise on the neutrino feature is then
\ba
\frac{S}{N} &\sim& f_\nu \left(1 + q \frac{b_0 -1}{b_0}\right)\sqrt{\bar{n}P_{hh}(k_{large})N_{k_{large}}}\\
&\sim& 0.1 b_0 \left( \frac{M_\nu}{0.2\eV}\right)  \left(1 + q \frac{b_0 -1}{b_0}\right) \sqrt{\bar{n}_hP_{hh}(k_{large}) \frac{V_{survey}}{(h^{-1} \Gpc)^3}\left(\frac{k_{large}}{0.01 h/\Mpc}\right)^3}
\ea
where $k_{large}$ is chosen to be just smaller than the neutrino free streaming scale $k_{fs}$.

For comparison, a similar estimate for the signal-to-noise for the scale-dependent bias due to primordial non-Gaussianity \cite{Dalal:2007cu, Seljak:2008xr,Ferraro:2014jba},  $\Delta b(k) = 2 f_{NL}(b_0-1)\delta_{crit}/\alpha(k)$ where $\alpha(k) = 2k^2T(k)D(z)/(3\Omega_m H_0^2)$,  is
\be
\frac{S}{N} \sim 0.05 f_{NL}(b_0 - 1) \left(\frac{0.001 h/\Mpc}{k_{large}}\right)^2 \sqrt{\bar{n}_hP_{hh}(k_{large}) \frac{V_{survey}}{(h^{-1} \Gpc)^3}\left(\frac{k_{large}}{0.001 h/\Mpc}\right)^3}\,.
\ee
So, detecting the neutrino scale-dependent bias with $M_\nu \sim 0.1\eV$ is comparable to detecting the non-Gaussian scale-dependent bias with $f_{NL} \sim 1$. 

The signal-to-noise on the neutrino feature in the ratio of the bias factors for two tracers with biases $b_1$, $b_2$ and number densities $\bar{n}_1$, $\bar{n}_2$ is 
\ba
\frac{S}{N} &\sim& q f_\nu \frac{b_2}{b_1} \left( \frac{b_1 - 1}{b_1} - \frac{b_2-1}{b_2}\right) \sqrt{\frac{P_{11}(k_{large}) N_{k_{large}}}{\frac{1}{\bar{n}_1}\frac{b_2^2}{b_1^2} + \frac{1}{\bar{n}_2} + \frac{1}{\bar{n}_1\bar{n}_2P_{11}}}}\\
&\sim &  0.1 \frac{b_2}{b_1}\left( \frac{M_\nu}{0.2\eV}\right) \left( \frac{b_1 - 1}{b_1} - \frac{b_2-1}{b_2}\right)\sqrt{\frac{\bar{n}_1P_{11}(k_{large}) \frac{V_{survey}}{(h^{-1} \Gpc)^3}\left(\frac{k_{large}}{0.01 h/\Mpc}\right)^3}{\frac{\bar{n}_1}{\bar{n}_2} + \frac{b_2^2}{b_1^2} + \frac{1}{\bar{n}_2P_{11}}}}
\ea
where $P_{11}$ is the autopower spectrum of galaxy population $1$.

On the other hand, the signal-to-noise on the $f_\nu$ dependence of $f(k)$, the derivative of the linear growth, is
\ba
\frac{S}{N}&\sim& \frac{3}{5}f_\nu \Omega_{m}^{6/11}(z)\sqrt{\frac{4}{45}\bar{n} P_{hh}(k) N_k}\\
&\sim& 0.6 \left(\frac{M_\nu}{0.2\eV}\right)\Omega_m^{6/11}(z) \sqrt{\bar{n}_hP_{hh}(k_{large}) \frac{V_{survey}}{(h^{-1} \Gpc)^3}\left(\frac{k_{max}}{0.1 h/\Mpc}\right)^3}\,.
\ea
In contrast to the scale-dependent bias, we have allowed information about $M_\nu$ from the amplitude of $f(k,z)$ alone (as opposed to just the scale dependence) so the signal-to-noise is dominated by information coming from the smallest scales (where $N_k$ is largest) and is therefore much larger. 

In summary, detecting the scale-dependent bias due to neutrino mass at the $\sim 0.1eV$ level  is comparable detecting the scale-dependent bias due to $f_{NL}\sim \mathcal{O}(1)$. The different scale-dependencies of the two signals, however, cause the $S/N$ to be dominated by different $k$. The optimal survey configurations for these signals are therefore likely different and the systematics limiting the two signals will be different as well. In particular, at large scales the non-Gaussian signal scales as $1/k^2$ so the signal-to-noise continues to increase with decreasing $k_{large}$. On the other hand, the neutrino feature is smooth change in amplitude between $k\ll k_{fs}$ and $k\gg k_{fs}$ so (with this estimate that assumes high $k$ are measured more precisely than low $k$) the $S/N$ is dominated by intermediate scales near $k_{fs}$ even if the minimum $k$ measured is smaller. 

\section{Data sets}
\label{sec:data}

\subsection{3D galaxy Power Spectrum from a Spectroscopic Survey}
The 3D power spectrum for sources with bias $b_g(k,z)$ is given by
\be
\label{eq:3DPower}
P_{gg}(k_{\perp},  k_{||}, z) = \left(b_g(k, z)+ f(k,z)\frac{k_{||}^2}{k^2}\right)^2P_{mm}(k)\,.
\ee
We treat different redshift bins as uncorrelated and model the covariance matrix of the 3D power spectra as
\be
C_{ij}(k, \mu, \bar{z}) = \sqrt{P_{gi}(k, \mu,\bar{z})P_{gj}(k, \mu,\bar{z})} + \frac{\delta_{ij}}{\bar{n}_i} 
\ee
where $gi$ and $gj$ refer to different galaxy samples with bias factors $b_{gi}$, $b_{gj}$ and number densities $\bar{n}_i$, $\bar{n}_j$. The Fisher matrix for the 3D power spectra is given by
\be
\label{eq:Fisher3D}
F_{\mathcal{OO}'} = \frac{1}{2}\sum_{\bar{z}}V(\bar{z})\int_{k_{min}}^{k_{max}}\int_{-1}^1 \frac{k^2 dkd\mu}{(2\pi)^2}{\rm Tr}\left(\frac{\partial C}{\partial \mathcal{O}}C^{-1}\frac{\partial C}{\partial \mathcal{O}'}C^{-1}\right)\,,
\ee
where $\bar{z}$ is the mean redshift, $V(\bar{z})$ is the volume, and we take $k_{min} = 0.001 h/\Mpc$ and $k_{max} = 0.1 h/\Mpc$  \cite{Tegmark:1996bz,Asorey2012,Dore:2014cca}. We have checked that including a prescription for bulk flows in Eq.~(\ref{eq:3DPower}) as in \cite{Dore:2014cca} does not significantly alter our results. 

\subsection{2D Galaxy and Lensing Spectra from a Broad-band Photometric Survey}

Now suppose we have a set of photometric measurements of the galaxy overdensity in direction specified by $\thetaB$ and in redshift bins defined by window function $W(z, z_s)$,
\be
\hat\delta_g(\thetaB, z_s) = \int dz W(z, z_s) \hat\delta_g(\chi(z)\thetaB)
\ee
where $\chi(z)$ is the comoving distance to redshift $z$. Further, suppose that we have a measurement of the lensing convergence from sources in each redshift bin, 
\be
\hat{\kappa}(\thetaB, z_s) = \int d\chi' g(\chi', \chi_s) \chi'\nabla_\perp^2\hat{\Phi}(\chi'\thetaB, z')\\
\ee
where
\be
g(\chi,\chi_s) \equiv \int_{z}^\infty dz'\frac{\chi'-\chi}{\chi'}W(z', z_s)
\ee 
where $\chi_s$ is the comoving distance to $z_s$ and $\chi'$ the comoving distance to $z'$ and $\nabla_\perp^2\hat\Phi$ is the Laplacian of the peculiar gravitational potential in the direction transverse to the line of sight.  

The angular power spectrum or cross-power spectrum for two populations galaxies in the same redshift bin is then
\ba
\label{eq:Cgg}
C_{\ell}^{g_ig_j}(z_g) & = & \frac{(4\pi)^2}{(2\pi)^3} \int k^2 dkP_{g_ig_j}(k, z_g) \int dz W(z, z_g) j_\ell(k\chi) \int dz' W(z', z_g) j_\ell(k\chi') \nn\\
&\approx& \int dz\, W^2(z, z_g) \frac{H(z)}{\chi^2} b_{g_i}b_{g_j} P_{mm}\left(\frac{\ell + 1/2}{\chi}, z\right)+\, \delta_{ij} s_i
\ea
where in the second line we have used the Limber approximation \cite{LoVerde:2008re}. The second term  $s_i$ is the shot noise, e.g.
\be
s_i =  \int dz\, W(z, z_g) \frac{H(z)}{\chi^2}\frac{1}{\bar{n}_i}\,.
\ee
The galaxy-convergence cross-power spectrum is given by
\be
\label{eq:Cgkappa}
C_\ell^{g\kappa}(z_g, z_s) 
\approx\frac{3}{2}H_0^2\Omega_m\int d\chi\, g(\chi, \chi_s) W(z, z_g)\frac{H(z)}{\chi}b_gP_{mm}\left(\frac{\ell+1/2}{\chi},z\right)(1+z)\,,
\ee
and the cross-power spectrum of the convergence field from sources at $z_1$ and $z_2$ is
\be
\label{eq:Ckappakappa}
C^{\kappa\kappa}_\ell(z_{s1},z_{s2})  
\approx \left(\frac{3}{2} \Omega_m H_0^2\right)^2\int d\chi' 
g(\chi', \chi_{s1})g(\chi', \chi_{s2})(1+z')^2P_{mm}\left(\frac{\ell + 1/2}{\chi'},z'\right)
 + \epsilon_\kappa
\ee
where $\epsilon_\kappa$ is the noise in the measurement of $\kappa$ (from e.g. shape noise) and in both Eq.~(\ref{eq:Cgkappa}) and Eq.~(\ref{eq:Ckappakappa}) we have used the Limber approximation. 

We will consider constraints from (i) the angular power spectrum of a single population of galaxies, (ii) the angular auto- and cross-power spectra of two populations of galaxies with different bias factors, and (iii) the angular auto and cross power spectra of a single population of galaxies and the lensing convergence. The Fisher matrix for the angular power spectra is 
\be
\label{eq:FisherCell}
F_{\mathcal{O}\mathcal{O'}} = \frac{f_{sky}}{2}\sum_{\ell} (2\ell + 1) Cov^{-1} \frac{dCov}{d\mathcal{O}}Cov^{-1}\frac{dCov}{d\mathcal{O}'}
\ee
where $Cov$ is the full covariance matrix of all of the angular auto and cross power spectra of the observables under consideration (e.g. the galaxy distribution and lensing convergence from each redshift bin).

\section{Forecasts}
\label{sec:forecast}
The expressions for the Fisher matrix  in Eq.~(\ref{eq:Fisher3D}) and Eq.~(\ref{eq:FisherCell}) can be used to study forecasted constraints on neutrino mass from the power spectra in Eqs.~(\ref{eq:3DPower}), (\ref{eq:Cgg}), (\ref{eq:Cgkappa}), (\ref{eq:Ckappakappa}). A central goal is to study the effects of scale-dependent halo bias $b(k)$ in Eq.~(\ref{eq:bk}) on the forecasted constraints on $M_\nu =  \sum_i m_{\nu i}$. A constant halo bias is, in principle, sensitive to neutrino mass through the dependence of the bias on the variance of mass fluctuations $\sigma^2(M)$, which is suppressed in cosmologies with massive neutrinos but only if one has an accurate model of halo bias in the first place. We will therefore consider the overall amplitude of the bias $b_0$ to be a free parameter without cosmological information and marginalize over it. To illustrate how the forecasted errors on $M_\nu$ depend on the bias $b_0$ and the number density $\bar{n}$ we will sometimes treat them as independent parameters. In the real universe the bias and number density are not independent, and to estimate $b_0(\bar{n})$ we use the mass function and bias expressions from \cite{Bhattacharya:2010wy} with the simplifying assumption that $n_{galaxy} = n_{halo}$. The galaxy population with the largest number density of tracers is assumed to have $b_0 = 0.8$ which is reasonable for a high-density faint sample. The bias of the second population of tracers is allowed to vary but see \cite{Zehavi:2004ii} for measurements of the bias factors for different galaxy populations. 

Since we are primarily interested in the improvement in constraints relative to the constant-bias case we keep all cosmological parameters aside from $\sum m_{\nu}$ and $A_s$ fixed.  We assume a cosmology with Hubble parameter $h = 0.67$, total (CDM + neutrino + baryon) matter density  $\Omega_m h^2 = 0.1419$, baryon density $\Omega_b h^2 = 0.022$, primordial amplitude of scalar perturbations $A_s = 2.215\times  10^{-9} (k/k_p)^{n_{s}}$, with $k_p = 0.05/\Mpc$ and $n_s = 0.96$. Forecasts are performed around a quasi-degenerate neutrino mass hierarchy with $m_{\nu i} = 0.1\eV$, $\sigma_8 = 0.82$ and when RSD information is included we marginalize over $\log_{10}A_s$. The publicly available CAMB code is used for all calculations of the transfer functions and power spectra \cite{Lewis:1999bs}. Throughout we restrict our forecast to large scales so that we may use the linear power spectra to calculate the observables in Eqs.~(\ref{eq:3DPower}), (\ref{eq:Cgg}), (\ref{eq:Cgkappa}), (\ref{eq:Ckappakappa}). 

For our broadband photometric survey we assume that the redshift distribution of all sources is given by
\be
\frac{dN}{dz} =  \frac{\beta z^\alpha}{\Gamma((\alpha+1)/\beta) z_0^{\alpha +1}}e^{-(z/z_0)^\beta}
\ee
with $z_0 = 0.57$, $\beta = 1.05$, and $\alpha = 1.26$ \cite{Chang:2013xja}. We assume that the galaxy population can be divided into two samples with different linear bias amplitudes given by $b_1(z) = 0.8D(z=0)/D(z)$ where $b_2(z) = 1.3 D(z=0)/D(z)$ where $D(z)$ is the linear growth factor.

The total angular density of sources across all redshifts is treated as a free parameter but we keep the ratio of the two populations fixed to $\bar{n}_1 = 4\bar{n}_2$. In forecasts that include lensing information, the effective number of sources for lensing is fixed to $\bar{n}_{eff} = \bar{n}_1/2$ \cite{Chang:2013xja}. 

The sources are further divided into redshift bins with ranges $z_{min}$ to $z_{max}$ defined by a window function,
\be
W(z| z_{min}, z_{max}) \propto \frac{dN}{dz}\left({\rm erfc}\left(\frac{z_{min} - z}{\sqrt{2}\sigma(z_{min})}\right) -{\rm erfc}\left(\frac{z_{max} - z}{\sqrt{2}\sigma(z_{max})}\right) \right)\,,
\ee
with $\sigma(z) = 0.01(1+z)$.  In the following, we take six redshift bins of width $\Delta z \approx  0.2$ defined by $z_{min}$ values $[0.05, 0.2, 0.4, 0.6, 0.8, 1.0]$ and corresponding $z_{max}$ values $[0.2, 0.4, 0.6, 0.8, 1.0, 1.2]$.

In forecasts using only measurements of the galaxy distribution (as opposed to those that include lensing information), we do not include $\sigma_8$ or $A_s$ because these parameters cannot be measured separately.  We treat the amplitude of the halo bias for each galaxy population and redshift bin as a free parameter and marginalize over it. Specifically we allow $N_{populations} \times N_{redshifts}$ independent halo biases, one for each redshift and galaxy population. In cases where we make a comparison to constraints from $P_{mm}(k)$, we have marginalized over the amplitude of $P_{mm}(k)$. Forecasts including information from the lensing convergence $\kappa$ marginalize over the amplitude $A_s$. 

\section{Results}
\label{sec:results}

\begin{figure}[t]
\begin{center}
$\begin{array}{cc}
\includegraphics[width=0.5\textwidth]{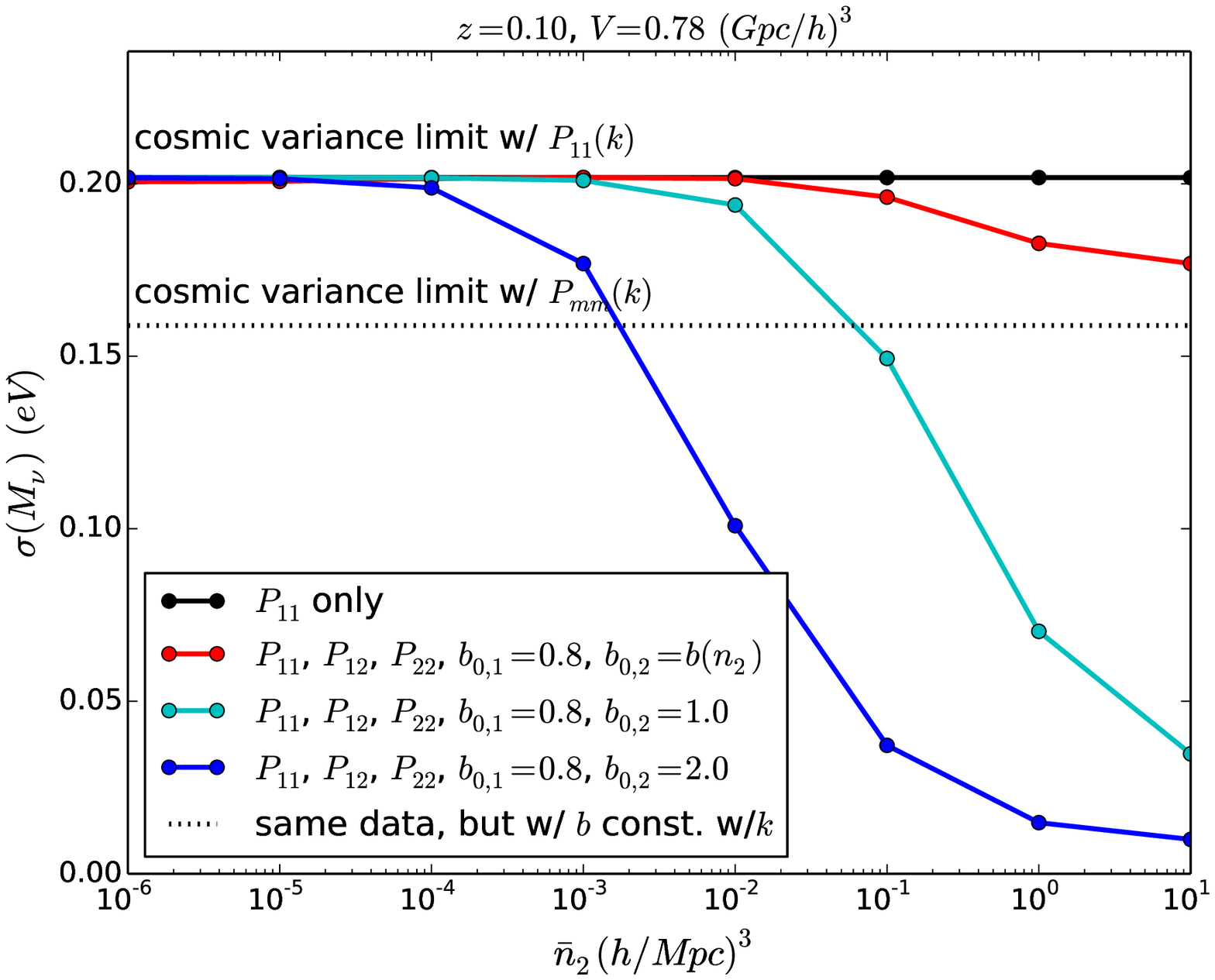} &  \includegraphics[width=0.5\textwidth]{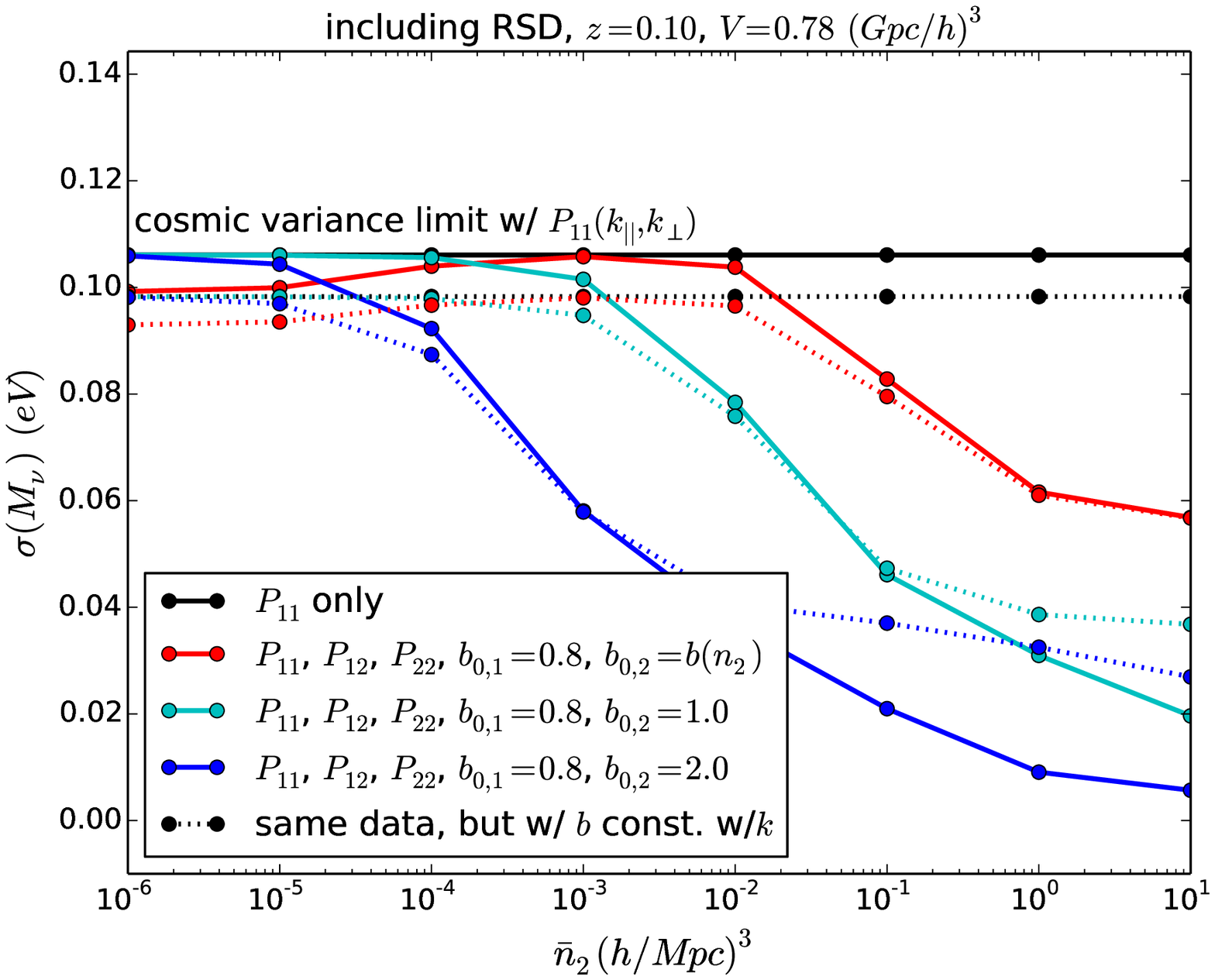} \\
\mbox{(a)} & \mbox{(b)}
\end{array}$
 \caption{\label{fig:3Derrors} Illustrations of the impact of using multiple tracers on the forecasted errors on $M_\nu$. Forecasts are for a measurement of the $3D$ auto- and cross-power spectra of two galaxy populations in a single redshift bin.  The left panel does not include information from redshift space distortions in the forecast the right panel includes information from the anisotropic power spectra including redshift space distortions and a $0.1\%$ prior on $\log_{10}A_s$. The number density of tracers of galaxy population $1$ is assumed to saturate the cosmic-variance limit for the assumed $k$-range (from $k_{min} = 10^{-3} h/\Mpc$ to $k_{max} = 0.1 h/\Mpc$), the number density of population $2$ is on the x-axis. The bias parameters $b_1$ and $b_2/b_1$ are marginalized over. The red curves with $b(n)$ use the mass functions and bias factors of \cite{Bhattacharya:2010wy} with the simplifying assumption $n_{galaxy} = n_{halo}$. Scale-dependent bias makes constraints on $M_\nu$ weaker, but if multiple tracers are used some of the information is recovered, and with a sufficient number of tracers the final constraints on $M_\nu$ are even stronger. Redshift space distortions considerably improve the constraints on $M_\nu$ and lessen the impact of scale-dependent bias on the final constraints until one reaches very large $n_2$.}
\end{center}
\end{figure}

The forecasted errors on $M_\nu$ from a single redshift bin, neglecting the redshift space distortion term $f(k,z)k_{||}^2/k^2$, are shown in the left panel of Figure \ref{fig:3Derrors}. We have chosen to plot $z=0.1$ because the largest number of tracers is available at low redshift. Increasing the number of redshift bins decreases the value of $\sigma(M_\nu)$ but does not change the trends with $\bar{n}_2$ or the relative positions of the curves on that graph.  If constraints come from the autopower spectrum of a single population of galaxies, the scale-dependent bias $b(k)$ degrades the constraints relative to forecasts (incorrectly) assuming a constant bias factor. This is to be expected since  scale-dependent bias introduces a degeneracy between $b_0$ and $M_\nu$. Furthermore, for sources with $b_{0} >1$, scale-dependent bias lessens the neutrino-induced suppression in $P_{g1g1}$ relative to $P_{mm}$. On the other hand, if multiple tracers of the density field are used then additional information about neutrino masses can be gained from the scale dependence of the ratio of the bias factors shown in Fig.~\ref{fig:bofk}. Beating cosmic variance, however, requires a huge number density of tracers and/or multiple tracers with very different bias factors. For example, one needs $\bar{n}\gsim 0.01 (h/\Mpc)^3$ for sources with $b_{0,2}/b_{0,1} = 1.25$  (see Fig.~\ref{fig:3Derrors} and \S\ref{ssec:estimates} for a quantitative presentation of the dependence of the errors on $b_{0,1}$, $b_{0,2}$ and $\bar{n}$).  On the other hand, if the redshift space information is included as in right panel of Fig. \ref{fig:3Derrors}, constraints on neutrino mass are always much stronger, the impact of scale-dependent bias is much weaker, and beating the cosmic-variance limit does not require such a high density of sources. For instance, for the same example of $b_{0,2}/b_{1,0} =1.25$, beating cosmic variance places a weaker requirement $\bar{n} \gsim 0.001 (h/\Mpc)^3$.

\begin{figure}[t]
\begin{center}
$\begin{array}{cc}
\includegraphics[width=0.5\textwidth]{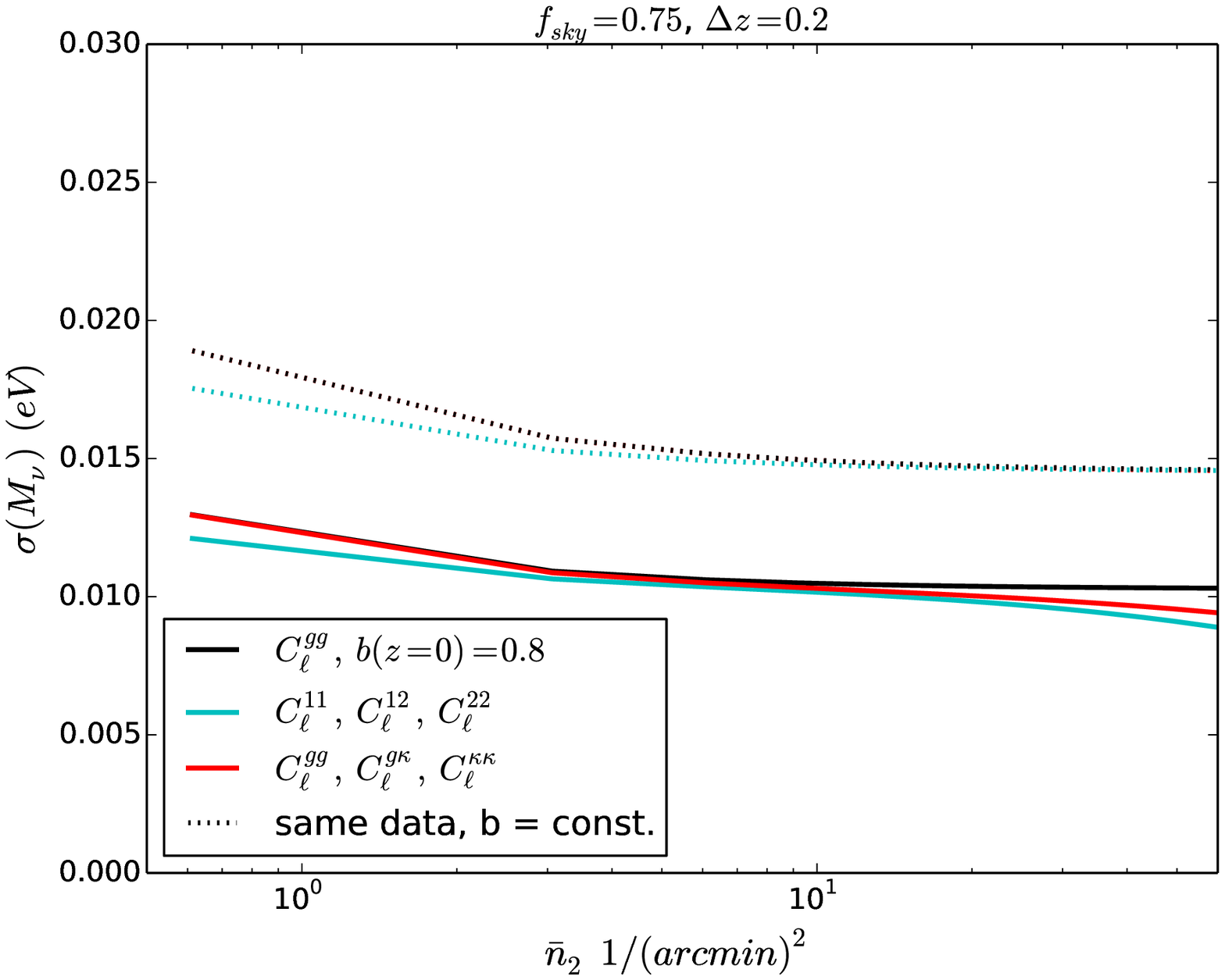} &\includegraphics[width=0.5\textwidth]{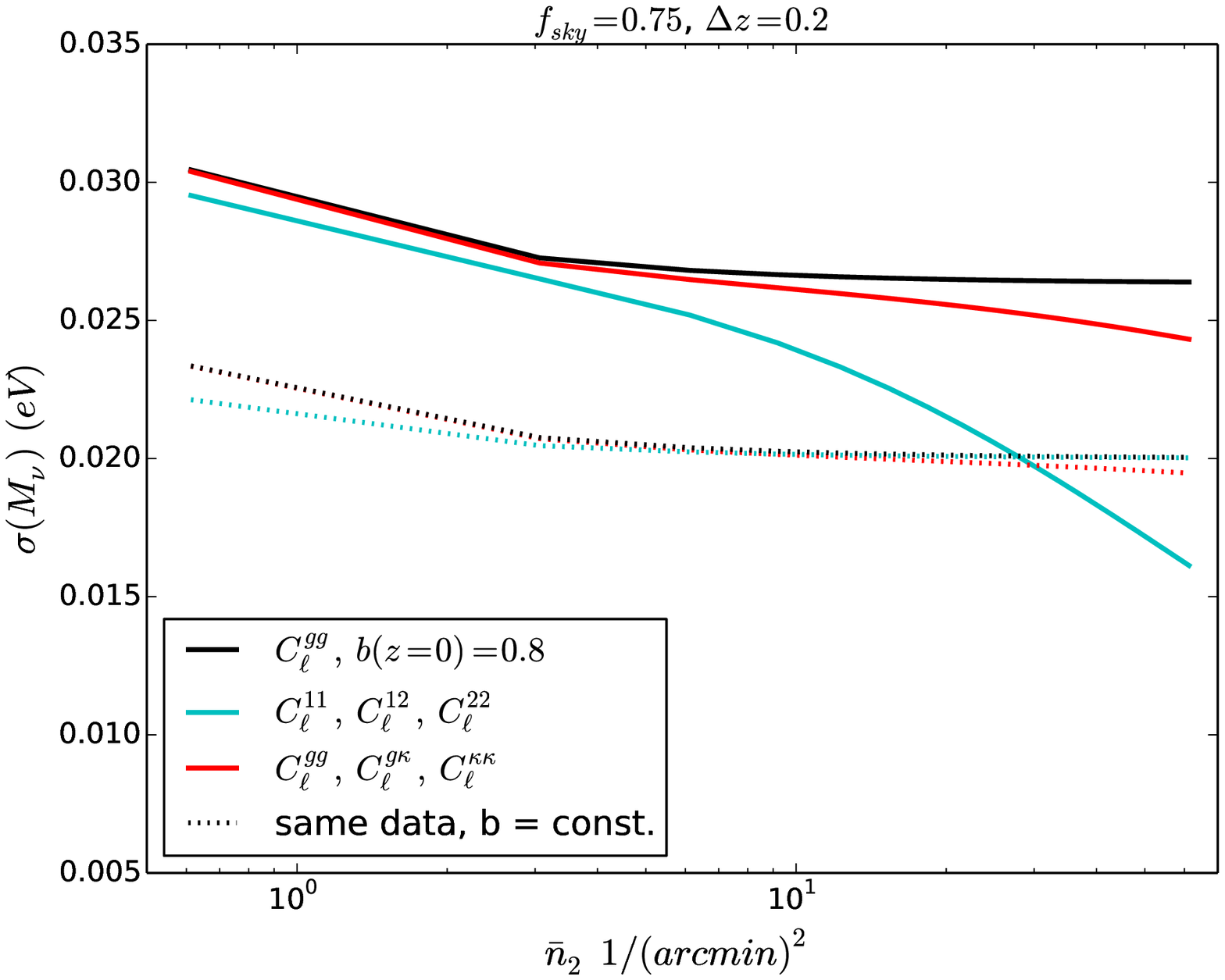} \\
\mbox{(a)} & \mbox{(b)}
\end{array}$
 \caption{\label{fig:2Derrors} Forecasted constraints on $M_\nu$ from the angular auto- and cross-power spectra of galaxies and the lensing convergence. The first population of galaxies has $\bar{n}_1 = 4 \bar{n}_2$, the number of sources for lensing convergence is taken to be $\bar{n}_{eff} = \bar{n}_1/2$ and we assume shape noise $\gamma = 0.16$ (see, e.g. \cite{Abbott:2005bi}). The left panel shows the unmarginalized constraints on $M_\nu$, while the right panel marginalizes over bias parameters for each sample in each redshift bin.  Scale-dependent bias causes a degeneracy between the bias parameters and the neutrino mass, significantly weakening constraints on $M_\nu$ in comparison with an incorrect model with $b = const.$. Using multiple tracers with different biases can bring dramatic improvements in $\sigma(M_\nu)$, ultimately beating the cosmic-variance limit on $M_\nu$ from the autopower spectrum of a single population of galaxies. }
\end{center}
\end{figure}

In Fig.~\ref{fig:2Derrors} the forecasted constraints on $M_\nu$ from measurements of the angular galaxy and lensing power spectra are shown.  For these forecasts, we assume a survey that covers $f_{sky}= 0.75$ and use the expression in Eq.~(\ref{eq:FisherCell}) summing from $\ell_{min} = 4$ to $\ell_{max} = 500$. The forecasted constraints, marginalizing over six independent bias parameters for each redshift bin, are shown in Fig.~\ref{fig:2Derrors}. For a single population of galaxies the forecasted errors on $M_\nu$ decrease with increasing $\bar{n}$ and then reach a plateau at the cosmic-variance limit. When the cross-power spectrum from either an additional sample of galaxies or the lensing convergence is included the errors continue to drop with increasing $\bar{n}$ eventually falling below the forecasted constraints for the $b= const.$ assumption.

\section{conclusions}
\label{sec:conclusions}
In this paper, we have shown that the scale dependence of the halo bias $b(k)$, or the scale-dependent ratio of the bias factors of two different populations $b_1(k)/b_2(k)$, provides a novel probe of the neutrino mass hierarchy. The linear bias, along with the logarithmic derivative of the linear growth function $f(k)$, can be measured without cosmic variance.  That is, the fundamental limit on the precision of these observables is not cosmic variance, but the number of tracers or the stochasticity between mass and galaxies. 

In practice, beating cosmic variance will be a challenge. The main limiting factor is the need for extremely high number densities of sources with different bias factors (see Fig.~\ref{fig:3Derrors},  Fig.~\ref{fig:2Derrors} and \ref{ssec:estimates}). Achieving such dense samples of galaxies is challenging. Additionally, the bias and number density are not independent parameters and the bias factors for very dense populations are generally not too different from $1$, but the amplitude of the neutrino feature in the ratio of the biases scales as $b_0-1$.  As a reference point, the densest spectroscopic sample of the planned DESI mission is expected to reach $\bar{n} \sim 0.005 (h/\Mpc)^3$ \cite{Levi:2013gra} while the highest-density sample of the proposed SPHEREx mission is expected to reach $\bar{n} \sim 0.01 (h/\Mpc)^3$ \cite{Dore:2014cca}. 

Another important result is that unless one has such extremely high-density samples, including the scale dependence of the bias in forecasts weakens constraints on $M_\nu$ from galaxy clustering because of the degeneracy between $b_0$ and $M_{\nu}$. While this is true for both angular power spectrum measurements (see Fig.~\ref{fig:2Derrors}) and 3D power spectrum measurements (see Fig.~\ref{fig:3Derrors}), the consequences of scale-dependent bias are much less severe for the 3D power spectra when information from redshift-space distortions is used. Scale-dependent bias, of course, has no impact on measurements of $M_\nu$ from lensing alone. 

While scale-dependent bias may not currently provide competitive constraints on neutrino mass, it is worth emphasizing that it is still a new observable that we should try to measure (see Figures \ref{fig:bofk} and \ref{fig:C12OC11}). In particular, the ratios of cross-power spectra to autopower spectra plotted in Fig. \ref{fig:C12OC11} are quantities that, without $b(k)$ would not depend on neutrino mass at all. Measuring these observables would provide a new confirmation of scale-dependent growth of perturbations in the $\nu$CDM universe. In this paper, we have restricted analysis to the large-scale linear bias factor and ignored possible complications due to nonlinear biasing or scale-dependent bias from other sources (see e.g. \cite{Biagetti:2014pha}). The presence of other sources of scale dependence would make it more difficult to interpret the signals in Fig.~\ref{fig:bofk} and Fig.~\ref{fig:C12OC11}. On the other hand, it may be possible to use the known shape and scaling of these signals with $b_0$ to extract information about $M_\nu$. It is also worth emphasizing that the neutrino feature in $b(k)$ is on the same physical scales as the neutrino suppression in the matter power spectrum, so systematic effects changing the shape of $b(k)$ on the scales of interest will already be systematics to measurements of $M_\nu$ from galaxy clustering.

\acknowledgements
M.L. acknowledges helpful discussions with Gary Bernstein, Neal Dalal, Wayne Hu, and especially Matias Zaldarriaga. M.L. is grateful to Olivier Dor\'{e} and Roland de Putter for helpful correspondence and comments on this draft, and for sharing the bias factors and redshift distributions of galaxy samples used for SPHEREx forecasts. This research was supported by the Department of Energy DE-FG02-13ER41958 and by  the National Science Foundation PHY1316617. 

\bibliographystyle{ieeetr}
\bibliography{mdm}

\begin{thebibliography}{10}

\bibitem{Ade:2015xua}
P.~A.~R. Ade {\em et~al.}, ``{Planck 2015 results. XIII. Cosmological
  parameters},'' 2015.

\bibitem{Agashe:2014kda}
K.~A. Olive {\em et~al.}, ``{Review of Particle Physics},'' {\em Chin. Phys.},
  vol.~C38, p.~090001, 2014.

\bibitem{Palanque-Delabrouille:2015pga}
N.~Palanque-Delabrouille {\em et~al.}, ``{Neutrino masses and cosmology with
  Lyman-alpha forest power spectrum},'' {\em JCAP}, vol.~1511, no.~11, p.~011,
  2015.

\bibitem{Cuesta:2015iho}
A.~J. Cuesta, V.~Niro, and L.~Verde, ``{Neutrino mass limits: robust
  information from the power spectrum of galaxy surveys},'' {\em Phys. Dark
  Univ.}, vol.~13, pp.~77--86, 2016.

\bibitem{Levi:2013gra}
M.~Levi {\em et~al.}, ``{The DESI Experiment, a whitepaper for Snowmass
  2013},'' 2013.

\bibitem{Euclid}
R.~{Laureijs}, J.~{Amiaux}, S.~{Arduini}, J.~. {Augu{\`e}res}, J.~{Brinchmann},
  R.~{Cole}, M.~{Cropper}, C.~{Dabin}, L.~{Duvet}, A.~{Ealet}, and et~al.,
  ``{Euclid Definition Study Report},'' {\em ArXiv e-prints}, Oct. 2011.

\bibitem{LSST}
{LSST Science Collaboration}, P.~A. {Abell}, J.~{Allison}, S.~F. {Anderson},
  J.~R. {Andrew}, J.~R.~P. {Angel}, L.~{Armus}, D.~{Arnett}, S.~J. {Asztalos},
  T.~S. {Axelrod}, and et~al., ``{LSST Science Book, Version 2.0},'' {\em ArXiv
  e-prints}, Dec. 2009.

\bibitem{WFIRST}
D.~{Spergel}, N.~{Gehrels}, C.~{Baltay}, D.~{Bennett}, J.~{Breckinridge},
  M.~{Donahue}, A.~{Dressler}, B.~S. {Gaudi}, T.~{Greene}, O.~{Guyon},
  C.~{Hirata}, J.~{Kalirai}, N.~J. {Kasdin}, B.~{Macintosh}, W.~{Moos},
  S.~{Perlmutter}, M.~{Postman}, B.~{Rauscher}, J.~{Rhodes}, Y.~{Wang},
  D.~{Weinberg}, D.~{Benford}, M.~{Hudson}, W.-S. {Jeong}, Y.~{Mellier},
  W.~{Traub}, T.~{Yamada}, P.~{Capak}, J.~{Colbert}, D.~{Masters}, M.~{Penny},
  D.~{Savransky}, D.~{Stern}, N.~{Zimmerman}, R.~{Barry}, L.~{Bartusek},
  K.~{Carpenter}, E.~{Cheng}, D.~{Content}, F.~{Dekens}, R.~{Demers},
  K.~{Grady}, C.~{Jackson}, G.~{Kuan}, J.~{Kruk}, M.~{Melton}, B.~{Nemati},
  B.~{Parvin}, I.~{Poberezhskiy}, C.~{Peddie}, J.~{Ruffa}, J.~K. {Wallace},
  A.~{Whipple}, E.~{Wollack}, and F.~{Zhao}, ``{Wide-Field InfrarRed Survey
  Telescope-Astrophysics Focused Telescope Assets WFIRST-AFTA 2015 Report},''
  {\em ArXiv e-prints}, Mar. 2015.

\bibitem{Dore:2014cca}
O.~DoreŽ, J.~Bock, P.~Capak, R.~de~Putter, T.~Eifler, {\em et~al.},
  ``{Cosmology with the SPHEREX All-Sky Spectral Survey},'' 2014.

\bibitem{Henderson:2015nzj}
S.~W. Henderson {\em et~al.}, ``{Advanced ACTPol Cryogenic Detector Arrays and
  Readout},'' pp.~1--8, 2016.

\bibitem{Benson:2014qhw}
B.~A. Benson {\em et~al.}, ``{SPT-3G: A Next-Generation Cosmic Microwave
  Background Polarization Experiment on the South Pole Telescope},'' {\em Proc.
  SPIE Int. Soc. Opt. Eng.}, vol.~9153, p.~91531P, 2014.

\bibitem{Abazajian:2013oma}
K.~N. Abazajian {\em et~al.}, ``{Neutrino Physics from the Cosmic Microwave
  Background and Large Scale Structure},'' {\em Astropart. Phys.}, vol.~63,
  pp.~66--80, 2015.

\bibitem{Allison:2015qca}
R.~Allison, P.~Caucal, E.~Calabrese, J.~Dunkley, and T.~Louis, ``{Towards a
  cosmological neutrino mass detection},'' 2015.

\bibitem{Manzotti:2015ozr}
A.~Manzotti, S.~Dodelson, and Y.~Park, ``{External priors for the next
  generation of CMB experiments},'' {\em Phys. Rev.}, vol.~D93, no.~6,
  p.~063009, 2016.

\bibitem{DiValentino:2015sam}
E.~Di~Valentino, E.~Giusarma, O.~Mena, A.~Melchiorri, and J.~Silk,
  ``{Cosmological limits on neutrino unknowns versus low redshift priors},''
  {\em Phys. Rev.}, vol.~D93, no.~8, p.~083527, 2016.

\bibitem{Lesgourgues:2006nd}
J.~Lesgourgues and S.~Pastor, ``{Massive neutrinos and cosmology},'' {\em
  Phys.Rept.}, vol.~429, pp.~307--379, 2006.

\bibitem{Hu:1997mj}
W.~Hu, D.~J. Eisenstein, and M.~Tegmark, ``{Weighing neutrinos with galaxy
  surveys},'' {\em Phys.Rev.Lett.}, vol.~80, pp.~5255--5258, 1998.

\bibitem{Font-Ribera:2013rwa}
A.~Font-Ribera, P.~McDonald, N.~Mostek, B.~A. Reid, H.-J. Seo, and A.~Slosar,
  ``{DESI and other dark energy experiments in the era of neutrino mass
  measurements},'' {\em JCAP}, vol.~1405, p.~023, 2014.

\bibitem{LoVerde:2014pxa}
M.~LoVerde, ``{Halo bias in mixed dark matter cosmologies},'' {\em Phys.Rev.},
  vol.~D90, no.~8, p.~083530, 2014.

\bibitem{Upadhye:2015lia}
A.~Upadhye, J.~Kwan, A.~Pope, K.~Heitmann, S.~Habib, H.~Finkel, and
  N.~Frontiere, ``{Redshift-space distortions in massive neutrino and evolving
  dark energy cosmologies},'' {\em Phys. Rev.}, vol.~D93, no.~6, p.~063515,
  2016.

\bibitem{Bernstein:2011ju}
G.~M. Bernstein and Y.-C. Cai, ``{Cosmology without cosmic variance},'' {\em
  Mon. Not. Roy. Astron. Soc.}, vol.~416, p.~3009, 2011.

\bibitem{McDonald:2008sh}
P.~McDonald and U.~Seljak, ``{How to measure redshift-space distortions without
  sample variance},'' {\em JCAP}, vol.~0910, p.~007, 2009.

\bibitem{Seljak:2008xr}
U.~Seljak, ``{Extracting primordial non-gaussianity without cosmic variance},''
  {\em Phys.Rev.Lett.}, vol.~102, p.~021302, 2009.

\bibitem{Hui:2007zh}
L.~Hui and K.~P. Parfrey, ``{The Evolution of Bias: Generalized},'' {\em
  Phys.Rev.}, vol.~D77, p.~043527, 2008.

\bibitem{Parfrey:2010uy}
K.~Parfrey, L.~Hui, and R.~K. Sheth, ``{Scale-dependent halo bias from
  scale-dependent growth},'' {\em Phys.Rev.}, vol.~D83, p.~063511, 2011.

\bibitem{Villaescusa-Navarro:2013pva}
F.~Villaescusa-Navarro, F.~Marulli, M.~Viel, E.~Branchini, E.~Castorina,
  E.~Sefusatti, and S.~Saito, ``{Cosmology with massive neutrinos I: towards a
  realistic modeling of the relation between matter, haloes and galaxies},''
  {\em JCAP}, vol.~1403, p.~011, 2014.

\bibitem{Castorina:2013wga}
E.~Castorina, E.~Sefusatti, R.~K. Sheth, F.~Villaescusa-Navarro, and M.~Viel,
  ``{Cosmology with massive neutrinos II: on the universality of the halo mass
  function and bias},'' {\em JCAP}, vol.~1402, p.~049, 2014.

\bibitem{Ringwald:2004np}
A.~Ringwald and Y.~Y. Wong, ``{Gravitational clustering of relic neutrinos and
  implications for their detection},'' {\em JCAP}, vol.~0412, p.~005, 2004.

\bibitem{LoVerde:2013lta}
M.~LoVerde and M.~Zaldarriaga, ``{Neutrino clustering around spherical dark
  matter halos},'' {\em Phys.Rev.}, vol.~D89, p.~063502, 2014.

\bibitem{LoVerde:2014rxa}
M.~LoVerde, ``{Spherical collapse in $\nu \Lambda$CDM},'' {\em Phys.Rev.},
  vol.~D90, no.~8, p.~083518, 2014.

\bibitem{Baldauf:2013hka}
T.~Baldauf, U.~Seljak, R.~E. Smith, N.~Hamaus, and V.~Desjacques, ``{Halo
  Stochasticity from Exclusion and non-linear Clustering},'' {\em Phys.Rev.},
  vol.~D88, p.~083507, 2013.

\bibitem{Patej:2015lwa}
A.~Patej and D.~Eisenstein, ``{Quantifying the Colour-Dependent Stochasticity
  of Large-Scale Structure},'' 2015.

\bibitem{Pujol:2015wna}
A.~Pujol, K.~Hoffmann, N.~Jiménez, and E.~Gaztañaga, ``{What determines large
  scale clustering: halo mass or environment?},'' 2015.

\bibitem{Wang:1998gt}
L.-M. Wang and P.~J. Steinhardt, ``{Cluster abundance constraints on
  quintessence models},'' {\em Astrophys.J.}, vol.~508, pp.~483--490, 1998.

\bibitem{Hu:1997vi}
W.~Hu and D.~J. Eisenstein, ``{Small scale perturbations in a general MDM
  cosmology},'' {\em Astrophys.J.}, vol.~498, p.~497, 1998.

\bibitem{Anderson:2013yy}
L.~Anderson {\em et~al.}, ``{The clustering of galaxies in the SDSS-III Baryon
  Oscillation Spectroscopic Survey: baryon acoustic oscillations in the Data
  Releases 10 and 11 Galaxy samples},'' {\em Mon. Not. Roy. Astron. Soc.},
  vol.~441, no.~1, pp.~24--62, 2014.

\bibitem{Dalal:2007cu}
N.~Dalal, O.~Dore, D.~Huterer, and A.~Shirokov, ``{The imprints of primordial
  non-gaussianities on large-scale structure: scale dependent bias and
  abundance of virialized objects},'' {\em Phys.Rev.}, vol.~D77, p.~123514,
  2008.

\bibitem{Ferraro:2014jba}
S.~Ferraro and K.~M. Smith, ``{Using large scale structure to measure $f_{NL},
  g_{NL}$ and $?_{NL}$},'' {\em Phys. Rev.}, vol.~D91, no.~4, p.~043506, 2015.

\bibitem{Tegmark:1996bz}
M.~Tegmark, A.~Taylor, and A.~Heavens, ``{Karhunen-Loeve eigenvalue problems in
  cosmology: How should we tackle large data sets?},'' {\em Astrophys.J.},
  vol.~480, p.~22, 1997.

\bibitem{Asorey2012}
J.~{Asorey}, M.~{Crocce}, E.~{Gazta{\~n}aga}, and A.~{Lewis}, ``{Recovering 3D
  clustering information with angular correlations},'' {\em
  Mon.Not.Roy.Astron.Soc.}, vol.~427, pp.~1891--1902, Dec. 2012.

\bibitem{LoVerde:2008re}
M.~LoVerde and N.~Afshordi, ``{Extended Limber Approximation},'' {\em
  Phys.Rev.}, vol.~D78, p.~123506, 2008.

\bibitem{Bhattacharya:2010wy}
S.~Bhattacharya, K.~Heitmann, M.~White, Z.~Lukic, C.~Wagner, {\em et~al.},
  ``{Mass Function Predictions Beyond LCDM},'' {\em Astrophys.J.}, vol.~732,
  p.~122, 2011.

\bibitem{Zehavi:2004ii}
I.~Zehavi {\em et~al.}, ``{The Luminosity and color dependence of the galaxy
  correlation function},'' {\em Astrophys. J.}, vol.~630, pp.~1--27, 2005.

\bibitem{Lewis:1999bs}
A.~Lewis, A.~Challinor, and A.~Lasenby, ``Efficient computation of {CMB}
  anisotropies in closed {FRW} models,'' {\em Astrophys. J.}, vol.~538,
  pp.~473--476, 2000.

\bibitem{Chang:2013xja}
C.~Chang, M.~Jarvis, B.~Jain, S.~M. Kahn, D.~Kirkby, A.~Connolly, S.~Krughoff,
  E.~Peng, and J.~R. Peterson, ``{The Effective Number Density of Galaxies for
  Weak Lensing Measurements in the LSST Project},'' {\em Mon. Not. Roy. Astron.
  Soc.}, vol.~434, p.~2121, 2013.

\bibitem{Abbott:2005bi}
T.~Abbott {\em et~al.}, ``{The dark energy survey},'' 2005.

\bibitem{Biagetti:2014pha}
M.~Biagetti, V.~Desjacques, A.~Kehagias, and A.~Riotto, ``{Nonlocal halo bias
  with and without massive neutrinos},'' {\em Phys. Rev.}, vol.~D90, no.~4,
  p.~045022, 2014.

\end{thebibliography}
\end{document}